\documentclass{ieeeaccess}
\usepackage{cite}
\usepackage{amsmath,amssymb,amsfonts}
\usepackage{algorithmic}
\usepackage{graphicx}
\usepackage{array}
\usepackage{pifont}
\newcommand{\cmark}{\ding{51}}%
\newcommand{\xmark}{\ding{55}}%
\usepackage{multirow}
\usepackage{placeins}
\usepackage{hyperref}
\usepackage{soul}
\graphicspath{{pict/}}
\usepackage{textcomp}
\def\BibTeX{{\rm B\kern-.05em{\sc i\kern-.025em b}\kern-.08em
    T\kern-.1667em\lower.7ex\hbox{E}\kern-.125emX}}
\begin{document}


\history{Received October 25, 2021, accepted November 21, 2021, date of publication November 25, 2021, date of current version December 3, 2021.}
\doi{10.1109/ACCESS.2021.3130741}

\title{An Outlook on the Future Marine Traffic Management System for Autonomous Ships}
\author{\uppercase{Michele Martelli}\authorrefmark{1},
\uppercase{Antonio Virdis}\authorrefmark{2},\uppercase{ Alberto Gotta}\authorrefmark{3},  \uppercase{Pietro Cassarà}\authorrefmark{3} and \uppercase{Maria Di Summa}\authorrefmark{4}}
\address[1]{Department of Naval Architecture, Electric, Electronic and Telecommunication Engineering (DITEN), University of Genoa, Genoa, Italy}
\address[2]{Department of Information Engineering, University of Pisa, Pisa, Italy}
\address[3]{Institute of Information Science and Technologies "Alessandro Faedo", National Research Council, Pisa, Italy}
\address[4]{Institute of Intelligent Industrial Technologies and Systems for Advanced Manufacturing, National Research Council, Bari, Italy}
\markboth
{Martelli \headeretal: An Outlook on the Future Marine Traffic Management System for Autonomous Ships}
{Martelli \headeretal: An Outlook on the Future Marine Traffic Management System for Autonomous Ships}
\corresp{Corresponding author: Michele Martelli (e-mail: michele.martelli@unige.it).}
\begin{abstract}
In the shipping digitalisation process, the peak will be reached with the advent of a wholly autonomous and at the same time safe and reliable ship. Full autonomy could be obtained by two linked Artificial-Intelligence systems representing the ship navigator and the ship engineer that possess sensing and analysis skills, situational awareness, planning, and control capabilities. Many efforts have been made in developing onboard systems; however, the shore facilities are not ready yet to deal with these new technologies. The paper aims to present the innovative technologies and methodologies needed to develop a futuristic Vessel Traffic System. The proposed systems will aim at faultless data acquisition and processing, provide input to decision-making systems, and suggest evasive manoeuvre; to deal with hazards and systems failure without human intervention onboard. The system is composed of three different and interacting layers. The first is an artificially intelligent tool to detect and control autonomous ships, thanks to situation recognition and obstacle avoidance strategies. The second is an orchestration and management platform designed to coordinate the sensing/actuation infrastructure and the AI algorithms' results made available by multiple ships, mustering edge, and distributed computing techniques to fulfil the specific harsh requirements of the sea environment. The final part is a holistic guidance-navigation-control framework to manage autonomous ships' navigation in a crowded area.
Eventually, a cyber-physical scenario, using both a ship digital-twin and a real model-scale ship, is suggested to test and validate the innovative system without the availability of a full-scale scenario.
\end{abstract}
\titlepgskip=-15pt
\maketitle
\section{Introduction}
\label{sec:intro}
\FloatBarrier
\PARstart{T}{he} growth of interest in Industry 4.0 affects all the industrial sectors, including the marine fields \cite{r4} and, in particular, the blue economy and efficient transportation.\\
This revolution leads to new challenging scenarios; most of the maritime research community's efforts and related stakeholders are now devoted to the autonomous ship. Shortly, we will see human-crewed and uncrewed vessels operating in the same sea area. Such a revolution needs a solid multidisciplinary scientific background, grouping capabilities from different fields.
There is still a considerable gap between the amount of data available and the ability of the marine sector to benefit from it. Deep learning algorithms represent the concrete possibility of elaborating this enormous amount of data into useful information to let ships navigate autonomously. Two of the most challenging research problems still open are the collision avoidance strategy and the coordination among different vessels \cite{r20}.\\
The challenge to which the maritime sector is called is integrating the new digitalised processes and technologies, mainly devoted to the intelligent exploitation of the amount of available data produced by sensors, some of those already deployed on most recent vessels.\\
The main marine stakeholders' goal is to achieve autonomous navigation as soon as possible, reduce costs and achieve safe, reliable, and efficient navigation \cite{r6}. Today what seemed impossible until a few years ago appears feasible thanks to the growing evolution of critical enabling technologies such as Artificial Intelligence, Big Data Analytics, and Virtual/Augmented Reality.\\
Based on the reasons above, new and technologically advanced vessel traffic systems are needed to populate the coasts worldwide since the current system \cite{r17,r23,r8} will be unable to deal with the new millennium challenges \cite{r13}.\\
Within this context, this paper looks for near-future and beyond near-future navigation scenarios, wherein ships with different degrees of autonomy \cite{r11}\cite{r41} will navigate the same area and proposes a new Vessel Traffic System architecture. Engineers will face a new challenge in managing this situation. The proposed layout is based on massive ICT technology use and the latest research in the marine engineering field.
In particular, it will be impossible to imagine that an autonomous ship can take some action without interacting with each other; for such a reason, advances in the Vessel Traffic System (VTS) design are necessary.

\section{Related Works}
\label{sec:rel}
In the last decade, Autonomous Surface Vessels (ASVs) have been the maritime community's main research topic.  Most of the efforts are focused on developing more reliable algorithms for autonomous navigation, guidance, and control \cite{r4}\cite{r40}. These algorithms can limit the human operator's errors, driving the waterborne transport towards more energy-efficiency, more safety, reducing the overall operating expenditure \cite{r1}.
Since maritime industry is showing the increasing interest in ASVs, the Maritime Safety Committee (MSC) of the International Maritime Organization (IMO) approved in June 2019 \cite{IMO} interim guidelines for Maritime Autonomous Surface Ships (MASS) trials.
These guidelines identifies four levels of autonomy degree which are:
\begin{itemize}
    \item Degree one: ships with automated processes and decision support, which includes the automation of some unsupervised operations but with a seafarer ready to take the control;
    \item Degree two: remotely controlled ships with seafarers on board, where the ships are operated from a remote location but the seafarers can be available on board to take the control;
    \item Degree three: Remotely controlled ship without seafarers on board; hence, the ship is controlled and operated from another location;
    \item Degree four: fully autonomous ship controlled by an operative system.
\end{itemize}
\subsection{Autonomous Navigation in the Shipping Sector}
The first technology review papers on ASVs were presented in \cite{r32}, where many universities and private entities' research projects are reported. In such a work, 60 prototypes of ASVs were classified according to the level of automation achieved.\\
An interesting study is shown in \cite{r36}, where authors analyse the results obtained during the ReVolt project, using a 1:20 scale model, to study advanced control systems to develop an unmanned ship navigation with zero-emission for the short sea shipping.\\
The trend analysis of the automation levels over the years, using the scale presented in \cite{schiaretti2017survey1} in the maritime sector, has been carried out, and the results are summarised in graphic format in  Figure \ref{fig:autonomy_trend}. It is worth noting that the number of autonomous vessels projects/prototypes increased over the years; in particular, the higher number is concentrated to a low level of autonomy because they could be of interest to the industry in a short time.\\
In \cite{r34},\cite{r35},\cite{r33}, and \cite{r18} the authors show the results of the experimental study conducted on the nonlinear control logic suitable for the manoeuvring operations of the autonomous ships. In \cite{r37} a new autonomous surface unit was designed. This unit can be configured as a large unit or as a fleet of smaller units that can transport autonomously one container \cite{r9}. \\
In \cite{r37}, the authors show the design of a new fleet of autonomous units, operating along inland waters for multiple purposes, such as passenger transport and the autonomous collection of floating waste.
In addition to research organisations; also the maritime industry, classification bodies and institutions, are putting effort into the realisation of autonomous vessels.\\
A crew-less ship needs communication and computing infrastructures that support intelligent algorithms/methods to implement automatic operation, navigation, and berthing. 

\Figure[t!](topskip=0pt, botskip=0pt, midskip=0pt)[width=0.9\linewidth]{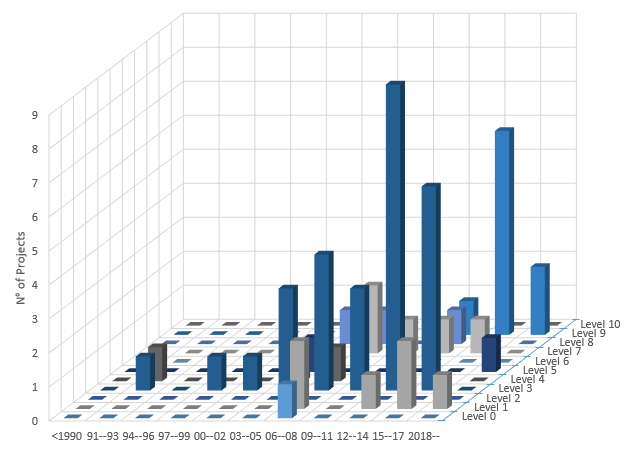}
{Trends over the year of unmanned project vs. degree of autonomy. \label{fig:autonomy_trend}}

\FloatBarrier

\subsection{Communication Technologies}
In \cite{r19}, the authors survey the leading Information and Communication Technologies (ICT), the communication architectures and some wireless standards proposed for ASVs. One of the emerging technologies to design communication infrastructure for this type of scenario is based on the Internet of Things (IoT) paradigm. In \cite{r31}, the authors highlight why IoT is generally considered the right candidate to support ships' efficient communication management. IoT technologies are now mature enough to be considered for the support of unmanned navigation systems as well. However, these systems need the support of intelligent algorithms and methods and an orchestration and management platform for handling the interactions with said intelligent algorithms, dynamically selecting the best level of automation for the ship. The complexity arising from IoT devices' interactions and the Cloud-computing environment is presented in \cite{r21}. The authors describe the challenges occurring in the specific maritime environment and propose possible approaches to tackle them. \\
Similarly, in \cite{r28} the authors describe how IoT technologies can be adopted in the naval industry, e.g., to facilitate services such as predictive maintenance, vessel tracking, and safety. Authors in \cite{r22} recognise the importance of ensuring the Quality of Services (QoS), safety, and security for sensor data and control data communications. For this purpose, they analyse the communication requirements coming from the various sensors involved in improving the ship's situational awareness, and they compare them against the different communication technologies. They also highlight the need for a connectivity manager to satisfy the various operating conditions' communication requirements. Work \cite{r10}, instead, investigates how protocols that are typically used in the IoT context can be applied to existing maritime communication technologies. For this purpose, they test on an experimental setup, a configuration using the Constrained Application Protocol (CoAP), i.e. a widely used IoT protocol, and IPv6 on top of a VHF radio technology. The obtained results prove the applicability of the considered protocols. \\
 In \cite{r50},{ the authors present the design framework of a Decision Support System (DSS) tool developed to assist the bridge operator during a challenging navigation condition. This work shows how using state of the art, IT devices and hardware, to increase safety at sea mainly focusing on both collision and grounding avoidance.}
\subsection{Computing Infrastructures}
As far as computing is concerned, all the intelligent applications based on AI technologies have to be managed, configured, and instantiated where needed, possibly ensuring a certain level of flexibility and dynamism to adapt to the ever-changing operation conditions. Moreover, said application needs to be logically interconnected with the IoT environment, e.g., discovering the available sensing nodes to gather all the information necessary for their functioning. To the best authors' knowledge, there is a substantial lack of maritime-specific solutions in this respect. In the context of Smart cities, several works discuss the problems related to obtain, maintain and expose a registry of the IoT devices that are available in the network, e.g. \cite{r15}, \cite{r14}, \cite{r25} and \cite{r16} The proposed solutions show that IoT registries can be efficiently maintained, aiming at optimising multiple requirements, such as energy efficiency, QoS, or security. Instead, in the aeronautical field, solutions exist to manage the dissemination of monitoring information in challenging environments, e.g. \cite{r2}.
\subsection{Application Services}
For what concerns the application management, the research has been lately focused on orchestrating applications in virtualised environments, e.g., either through Virtual Machines or following lightweight-virtualisation approaches such as Containers \cite{r29}. In this respect, the literature offers various methods to efficiently managing the life-cycle of IoT-based applications, e.g. \cite{r30} and \cite{r1}, which either consider allocating the execution of the applications in the cloud, i.e. remotely or at the edge of the network, thus potentially lowering communication delays. More recent works tackle the mobility of nodes and migrate applications as the user moves, trying to maintain proximity. These approaches typically are based on container migration techniques \cite{r7} and can minimise service disruption in all the application life-cycle.\\
In the building process of the new generation of ASVs, the last link in the chain is Artificial Intelligence (AI) and Augmented reality, which suitably combined can be used to design reliable, efficient, and comfortable ASVs. 
The synergistic development of the technologies mentioned in this work allows supporting the user in complex and challenging tasks. In \cite{Martelli_M} a decision support system with an augmented reality visualization system is presented.
The development process of UAV systems sometimes needs an amount of data not ever available, as illustrated in \cite{Alvey_2021_ICCV}; the authors overcome this bottleneck by introducing photorealistic data to augment the databases.
In other cases, the problem is the opposite. There are an enormous amount of data, and the available machines are limited, which is why \cite{Wang2020AugmentedRF} they propose a deep-learning algorithm that works in two phases, thus reducing the computational load of the single-stage.
The development of the trend towards a synergistic use leads to the need to have a unified interface between the various technologies used. The review \cite{MING202114} analyses the various ways of estimating the monocular depth, which is an important issue in the interface between deep learning and augmented reality.
Augmented reality is a fundamental tool for integrating information that the user already has in particularly complex or risky situations. The study described in \cite{jmse9090996} demonstrates the validity of supporting an augmented reality application for seven sailors involved in the activities of an icebreaker.
For example, in \cite{r12} and \cite{r38}, authors' investigate artificial intelligence algorithms to support the operator in preventing collisions using simulated data.
Safety and the ability to prevent accidents represent fundamental aspects in navigation, and they are two of the topics on which various artificial intelligence methods are questioned. In \cite{9361551}, a hybrid system that uses fuzzy logic and ontologies is presented. The ontologies allow to build a wide knowledge base and within this it is possible to connect events with different descriptions. The fuzzy logic is used to calculate in real-time the probability of accidents\cite{SENOL201670}. So it is possible fill the gap in numerical results by its semantic descriptions, embedding fuzzy fault tree analysis in ontology-based tree structures. 
The use of deep-learning techniques guarantees a high level of reliability, which is why in areas such as aeronautics, it is increasingly used and experimented with even in the construction phases of aircraft \cite{9160183}.
As illustrated in \cite{bhattarai2020embedded}, the possibility of processing and displaying a massive amount of data in particularly hostile conditions such as during a fire favors or improves safety conditions.
Deep-learning algorithms together with digital visualization techniques of reality are used in many applications. To achieve the safety standards currently required of civil structures, the authors propose a monitoring system based on a digital model that constantly exchanges data with the real one \cite{9547763}.
As the review \cite{
app11020821} demonstrates, the problem of safety is fundamental in various sectors; that is why, given the enormous reliability of deep-learning systems, they demonstrate increasingly performing results in the most variegate applications.
The use of deep learning algorithms has presented fascinating results, especially for autonomous driving \cite{r27}. They guarantee decision support modules localised on board but part of a distributed navigation strategy \cite{r26}. 
In the field of autonomous driving, the idea of creating a digital driver that can interface both with other machines and with human users is particularly challenging today \cite{Fernndez2020AssociatedRA}.
The spread of deep learning in various sectors has also involved the maritime industry in recent years.
In fact, the authors of the review \cite{r24} discuss the leading deep learning methods explored in autonomous navigation. The paper highlights the advantages that this new technology offers over traditional machine learning techniques. Also, as reported in \cite{r42}, marine traffic management significantly benefits from using deep learning techniques.

Although the literature presents outstanding works on ICT architectures to support the development of ASVs, the rapid evolution of new paradigms of both communication and orchestration of computing resources require further studies on how to integrate and decline such emerging technologies in the context of the autonomous navigation.
To this end, this paper aims at presenting the integration of the innovative technologies and methodologies needed to develop a futuristic ASV.
To face this challenge, a novel architecture is proposed hereafter, which is founded on four pillars that are: navigation control, computing orchestration, communication and networking, and data analysis. The presented architecture is designed to scale according to the degree of autonomy that the system can undergo, according to the classes defined by IMO in \cite{IMO}.
To the best of authors' knowledge, this is the first proposal that faces such a multidisciplinary challenge at those different tiers.
The paper is structured as follows: Sec. \ref{sec:system} deals with the overall system description and the orchestration is shown in Sec. \ref{sec:iot}. The use of Intelligent algorithms have been described in Sec. \ref{sec:ai} while the communication infrastructure is shown in Sec. \ref{sec:comm}. The proposed benchmark to test the new vessel traffic management system is reported in Sec \ref{sec:bench}. Eventually, the conclusions are drawn with the expected results and potential benefits.

\FloatBarrier

\section{System Description}
\label{sec:system}
 \FloatBarrier
The navigation in the coastal area needs to be supervised by an Ashore Control Center (ACC) that manages a massive amount of data, receive, elaborate, and return navigation strategies to each ship in the area. However, there can be conditions wherein no ACC is available to supervise ships' navigation, either autonomous or manned, occurring in the same sea area, thus resulting in a more challenging scenario. To fit such a set of different situations, we design a system that is represented in Fig. \ref{fig:system_layout} and includes the following entities: ships, sensors, AI algorithms and ACC. The sensors deployed in the navigation area will provide useful information to the Ships, AI algorithms, and the ACC. Based on this information, the ship controller will plan or maintain the route to be followed. The AI algorithms will enrich the data from sensors to support the ship's decisions and the monitoring at the ACC. Finally, the ACC merges data coming from sensors, from the vessels and the AI, to supervise and coordinate the traffic in the area of its competence. Moreover, as a failure backup strategy, ACC operators are equipped with an augmented-reality device, and they can monitor the situation and plan any manual interventions.\\
The sensors installed on either ship or in the surrounding environment will be implemented following the Internet of Things paradigm to be easily accessible from all the involved entities.\\
Artificial Intelligence (AI) algorithms are needed to create an Intelligent System (IS) and improve the autonomous decision process.  They can be deployed and executed either or both on the ships and at the ACC. This system improves situational awareness through the identification of obstacles and their possible classification. The considerable amount of data coming from multiple and diverse sources, acquired by heterogeneous sensors, enables the analysis and extraction of valuable information for monitoring specific situations. The IS will use this data to aid the navigation from two perspectives: on one side, it extract new knowledge usable through augmented reality by the operators of the ashore control station dealing with the monitoring; on the other side, the obtained knowledge can be used onboard to allow the ship to carry out evasive manoeuvres and strategies in complete autonomy.\\
The control and supervisor systems layout have been composed of two layers interacting each other. The remote one aims to coordinate the traffic, and the local one needed to handle the single ship. For this purpose, a Guidance-Navigation-Control (GNC) system for autonomous navigation needs to be ready to receive external information. In particular, GNC using the results from the AI algorithms, can detect an obstacle (either fixed or moving), elaborate an optimum evasive route, and, thanks to its control system, can actuate the new route safely and accurately. 
However, in the case of a failure on any single component of the VTS, the collision avoidance is still active thanks to the local Guidance-Navigation-Control system. This solution will be clearly sub-optimal, since the coordination part with the other entities of the scenario will be missing. \\
Eventually, to support the operations in such a heterogeneous set of scenarios, an Orchestration and Communication Platform (OCP) will be developed to efficiently handle the interactions among the ISs and the GNCs residing on the same ship, on different ships, or also within the ashore control centre. The OCP will provide flexibility to the overall system, allowing each ship for dealing efficiently with each scenario that can occur. For example, the vessel will discover whether other autonomous ships are navigating in the same area or if an ACC is available to monitor and support the traffic.\\
To achieve the proposed goals, the VTS needs to operate on three main aspects: autonomous maritime navigation, the analysis of a massive amount of data through AI, and the orchestration and management of their interactions.\\
From the marine-engineering perspective, the main changes is a new layer that communicates with the AI algorithms, receives the raw data from sensors, and detects a potential collision. In case a collision risk is detected, an optimised routing algorithm provides an evasive route. The optimisation problem is highly constrained and needs to consider the real manoeuvrability ships' capability, weather conditions, propulsion plant and manoeuvring-device limits. Once a new route is assessed by employing way-points (two spatial coordinates plus time), coordinated and shared with the VTS, this information will be sent to the track-keeping module, who will perform the proper set-points to reach the way-point accurately and safely, as shown in Fig. \ref{fig:system_layout}.\\
From the data analysis perspective, the VTS uses AI techniques to integrate and analyse various heterogeneous data, including raw measurements and images. The AI algorithms are used to extract the information content about obstacles or ships, mine the correlations between them, and build models that would enable predicting the route evolution. This produced rich data that support both the GNC in executing its tasks and the ashore operators in their monitoring activity. The process will be fully automated; however, the information will be displayed at the ACC's human-operators using Augmented Reality techniques in case of necessity. The current situation is continuously updated, and real-time monitoring will be possible. Thus, identifying obstacles and ships will allow the motion-planning system to carry out appropriate manoeuvres and predict evasive routes.

\Figure[t!](topskip=0pt, botskip=0pt, midskip=0pt)[width=0.9\linewidth]{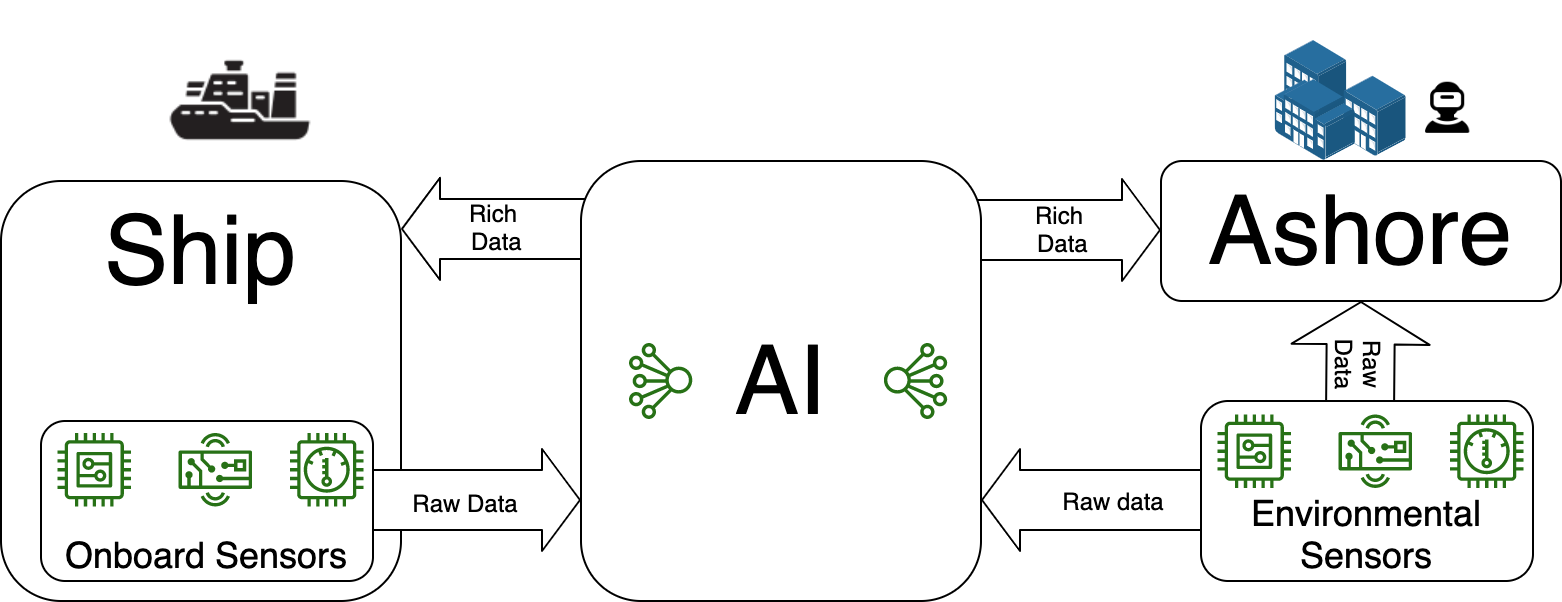}
{Layout of the marine traffic management. \label{fig:system_layout}}

\FloatBarrier

\section{IOT ORCHESTRATION}
\label{sec:iot}
 \FloatBarrier
The ISs used for situation awareness require a pervasive deployment of sensors and actuators to perform their tasks and aid the work of the GNCs. This will include sensors such as inertial motion units to monitor position, velocity, and acceleration; LiDAR and cameras; all the propulsion and manoeuvring monitoring devices; sonars and radars; etc. Moreover, they will need to interact with the GNCs to generate the proper set-point to control the machinery's behaviour. \\
While the GNCs will be deployed and interact only with the local physical environment, the ISs could either work only locally or possibly interact with remote entities, such as other ships or the ACC, when available. Moreover, to allow a flexible and cost-effective deployment, the GNCs should dynamically discover the ISs for decision support, available either locally or remotely, and to adapt to the context.\\
Fig. \ref{fig:iot_layout} shows an exemplified scenario wherein three ships, having diverse capabilities, coexist in the same area. A fourth node, the ACC, can also be considered if available in the scenario. Node A and C are ships equipped with an autonomous GNC and can execute Deep Learning algorithms for situation awareness. Node B is instead a crewed ship, which can measure and store environmental information.
The two above subsystems will need to be interconnected through an intelligent middleware that will ensure low latency in the monitoring-decision-actuation loop to allow correct and efficient operations. For this purpose, OCP has been defined; it manages the ISs life cycle, discovers and maintains a registry of the available sensing/actuation nodes, and schedules the communications among the various subsystems. The ability to grant the last requirement will also depend on the system's ability to adapt to the system's dynamic conditions, e.g. varying distance of the ship w.r.t. the ashore control centre, the varying performance of the available communication interfaces, etc. \\
The OCP acts as an intermediary between the ISs and the GNCs, providing the following operations. It will need to maintain a global vision of the system status, including the running AI applications status, together with a directory of the available sensing/actuation nodes. It will also need to know the Quality-of-Service requirements of each IS and/or control process within the GNCs to ensure said requirements are met during the execution. Whenever a new AI application is activated, the OCP needs to configure the system at multiple levels, operating in a cross-layer manner to let AI interact with the sensing entities available locally and possibly remotely. OCP allocates communication resources among the applications and said entities to fit the QoS requirements, e.g. communication latency below a certain threshold, a minimum granted data-transmission speed. Ensuring said QoS requirements are of paramount importance, especially when dealing with safety-critical operations, wherein low response times have to be granted either in nominal or contingency conditions, e.g., a congested communication network.

\Figure[t!](topskip=0pt, botskip=0pt, midskip=0pt)[width=0.9\linewidth]{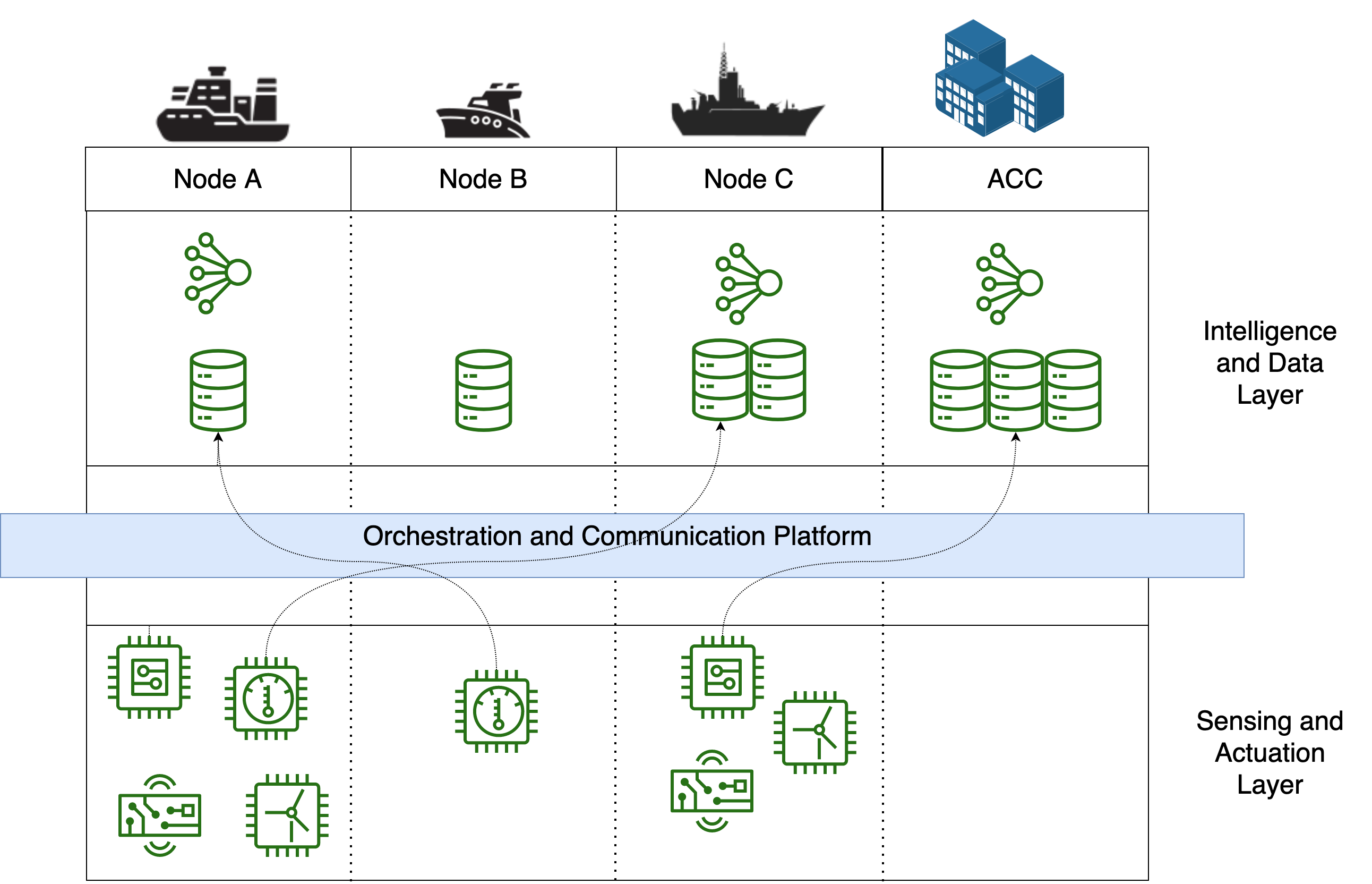}
{Logical system view.\label{fig:iot_layout}}

\FloatBarrier

\section{ARTIFICIAL INTELLIGENT ALGORITHMS AND AUGMENTED REALITY}
\label{sec:ai}
\FloatBarrier
The scientific idea outlined so far consists of a synergy of innovative technologies to implement a complex and highly automated system in which the coexistence of completely autonomous systems with systems under direct human control is possible. \\
Artificial intelligence is the ability of a machine to perform complex reasoning in a human-like way.
People often speak indifferently about Artificial Intelligence, Machine Learning, and Deep Learning. Indeed a distinction is necessary; Artificial Intelligence is a broad concept that includes automatic reasoning (machine learning), which provides a series of possible approaches, including deep learning. 
The deep learning-oriented approach to perform automated reasoning is the suggested path to fulfil the requirements to manage the traffic accounting for collision avoidance purposes. 
Deep learning algorithms enable the machine to autonomously classify data and structure them hierarchically. This allows the system to solve complex problems and gradually improve performance in an iterative learning process similar to that of the human mind.\\
A particularly lively current trend in artificial intelligence linked to learning services and communication efficient federated learning, is represented by auto-encoders. Auto-encoders are among the unsupervised neural networks. These networks can learn coded patterns of data, and subsequently, when faced with coded representations of data, they can re-generate the input data in an extremely accurate manner.\\
The fields of application that see a constant growth in the use of auto-encoders are the identification of anomalies and image de-noising. When declining the use of these particular networks in the marine field, an interesting prospect would be to use the auto-encoders to detect anomalies in the functioning of the onboard systems or reduce noise in the images of objects or ships coexisting in the same area of navigation. 
The idea is to use these networks to reconstruct the images of a ship or other object present in the navigation space, in conditions of poor visibility, where recognition even by a human operator is complex. 

The images are collected by cameras and then onboard-processed thanks to the auto-encoders, which allow extracting robust characteristics and recognizing boats or objects on the radar. Therefore, in addition to a detection problem, there is a problem of images registration from different cameras because the processed images are also different views of the same object. To solve these problems has been proposed the use of different types of auto-encoders to understand the best solution for each of the problems. The auto-encoders, as mentioned above, are designed to transform the input data into a latent code, that is, a compressed version of the starting image in which the features characterizing the object are preserved. Subsequently, this latent code is used to retrieve the relevant information. Below are some types of auto-encoders widely used and among which the authors believe to identify the best performing solutions for the proposed scenarios.
Under-complete auto-encoders are designed so that the latent code is presented with a lower dimensionality than the input. So if the input characteristics were all independent it would be extremely complex to reduce the dimensionality.
Denoising auto-encoders use images partially corrupted by the artificial and random addition of noise in the training phase to make the algorithm capable of recognizing and subtracting the noise and reconstructing the original data.
In sparse auto-encoders the dimensionality is not reduced, as in the previous cases by decreasing the neurons in the latent code, but the simultaneously active neurons are limited in number, so each input is represented by a limited number of neurons.
Variational auto-encoders are generative models, that is, after the training phase, they are able to generate images similar to those received in input. Therefore, they are able to extract fundamental features in the generation of latent code (reduced dimensionality) and then to generate images similar to the original ones.

The ability to detect anomalies could have various application scenarios, from identifying anomalies in the behaviour of other ships to the identification of anomalies in devices of a target ship. Another interesting hypothesis is represented by the use of auto-encoders for the extraction of features in hybrid structures to optimise already highly performing systems.\\
Instead, augmented reality is a technology that allows enriching the reality perceived by the human eye. Augmented Reality (AR) is a technology that increases the information content of the perceived reality of the user with additional digital objects and provides the user who uses it with a cognitive aid, but also with decision-support if digital objects are the result of an elaboration/processing of data and not a simple visualisation. AR is the optimal way to increase the situational awareness of the ACC operators, in case of emergency, when need to command ships remotely.
The goal is to train a system; using deep learning algorithms to recognise various situations; downstream of this recognition phase, the system provides the most suitable strategy to prevent a collision. Moreover, the most appropriate way to give the operator's processed information using virtual/augmented reality is considered. \\
The artificial intelligence system based on deep learning algorithms allows thanks to the analysis of a considerable amount of heterogeneous data (data from onboard sensors, images, information relating to the context in which the boat is operating, etc.) to perform an extremely accurate object recognition even in non-optimal conditions of poor visibility for example. In Fig. \ref{fig:ai} it is possible to see the intelligent system layout. Specific deep learning algorithms process heterogeneous data coming from multiple sensors, the outputs are suggestions to support operators and solutions for the autonomous ships. \\
The idea is to use artificial intelligence algorithms to support the operator in evaluating and choosing the proper operations to implement based on an analysis that considers as much data as possible. Operators will benefit from the intelligent processing of this data thanks to augmented reality.
Deep learning algorithms will allow training a system to recognise various situations thanks to the machine's ability to analyse a considerable amount of apparently unconnected data and then evaluate different solutions from time to time in a completely automatic way. As shown in Fig. \ref{fig:system_layout}, the intelligent system has the role of analysing and processing heterogeneous data coming from sensors but also from other ships, in fact it receives the data coming from every single ship and from the sensors installed in the surrounding environment, elaborate them, create and send the synthesis of command information to the GNC system of each system and display the information on the ACC (both on console and AR devices).\\
The use of innovative methodologies and strategies for data analysis to obtain an ever-increasing number of information allows interpreting the navigation scene sufficiently in advance to implement adequate preventive manoeuvres during navigation.
The goal is to use heterogeneous data to generate new relevant information relating to the acquired scene and interpret, through recognising particular objects, their classification, and the diagnosis of any interesting situations with deep learning techniques.
Another objective is to integrate the supporting digital information, not necessarily present in reality observed, to enrich the reference application context's reference case history.

\Figure[t!](topskip=0pt, botskip=0pt, midskip=0pt)[width=0.9\linewidth]{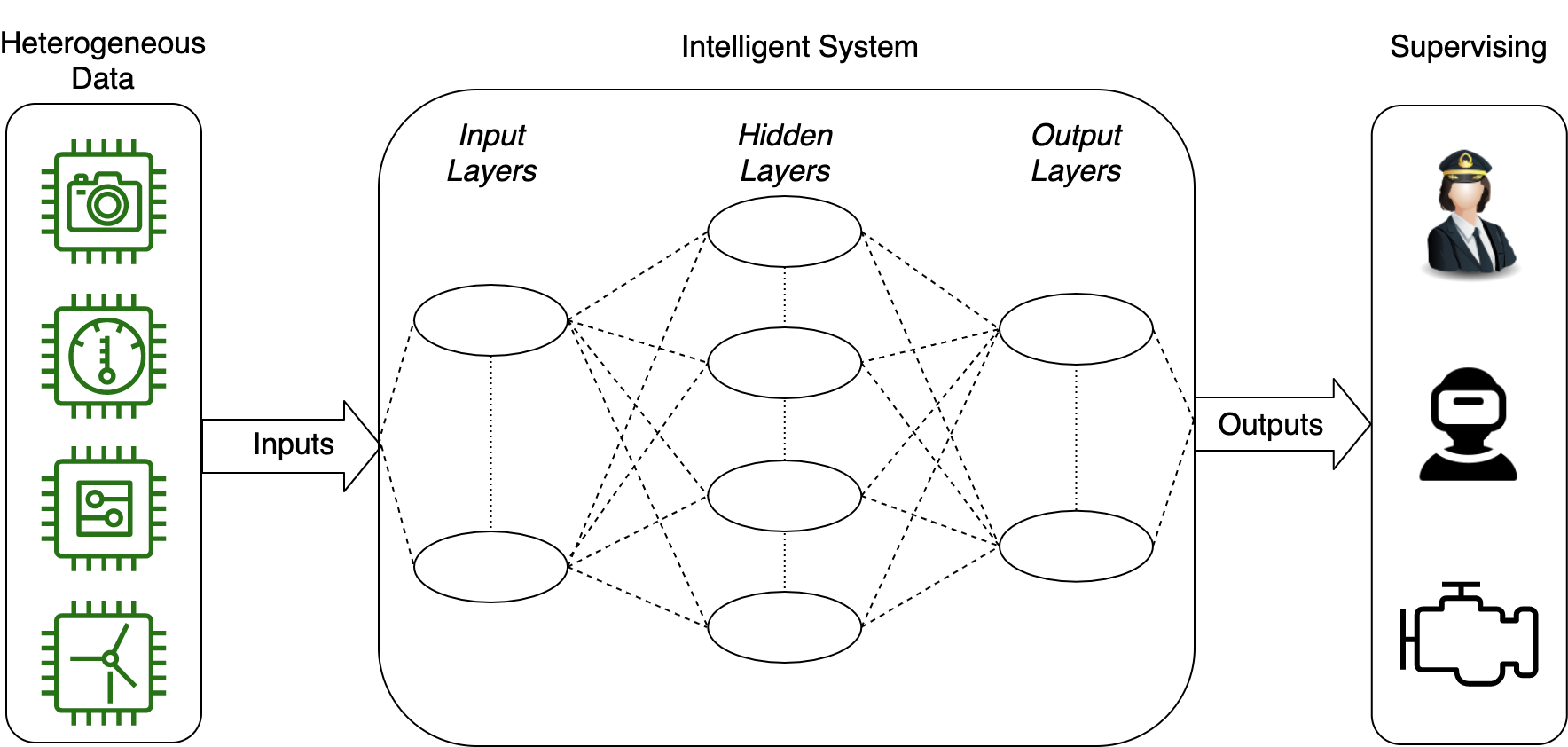}
{High-level view of the Intelligent System.\label{fig:ai}}
\FloatBarrier

\section{COMMUNICATIONS ARCHITECTURE}
\label{sec:comm}

\FloatBarrier

This section provides a review of the enabling technologies for maritime communications. These technologies can be used for the migration of current information communication systems towards more advanced systems based on a distributed communication logic, using, for example, Machine-Type Communication (MTC) and, in particular, the IoT paradigm.\\
The command and control of the ASVs are based on reliable algorithms for autonomous navigation and guidance. The use of these kinds of algorithms reduces the need for human operator's interaction with the vessel and increases exponentially the amount of data exchanged between ships or between a vessel and the ACC. In scenarios where the human component is limited to the control systems' detriment, MTCs are the most appropriate technological solution to support the communications. In particular, performing the control through thousands of sensors disseminated within the vessel system, the IoT paradigm \cite{r43}, \cite{r44}, and \cite{r45} is the most appropriate to deploy network architectures suitable for different application scenarios for autonomous shipping, such as search and rescue, aids-to-navigation and smart navigation. From the examples of autonomous shipping presented above, it is evident that the fundamental goal for network infrastructure is to provide reliable connectivity to heterogeneous types of systems and maritime applications and services, potentially based on the IoT communication paradigm.  The maritime scenario's challenges need to be addressed and modify the communication infrastructures accordingly to provide reliable network infrastructure.\\
The network infrastructure has to provide ubiquitous connectivity between vessels and ashore stations, on a global scale, especially over open oceans, to ensure the unbroken and consistent existence of services, such as those based on AI algorithms used for the assisted-navigation. Note that the ubiquitous connectivity needs of the roaming service among countries traversed by the vessels during navigation. In this scenario, the network infrastructure must be based on communication protocols that can convey nonuniform data traffic. Indeed, marine traffic near the ports, shore stations, and waterways is very dense. Quite the contrary, marine traffic on the high seas, mainly generated from intercontinental transportation or blue water shipping, is relatively sparse in density. Due to maritime services' multiplicity, it is needed to design the network infrastructure adopting a Service-oriented logic. Maritime applications and services vary from simple periodic reporting to route exchange and remote control, such as in the autonomous shipping use case.  So the communication infrastructure will be required to support a wide variety of maritime services, adapting itself to specific needs and match changing resource demands of those services. Hence both network configuration and communication resources must be made flexible and adaptive to the offered service. \\
Further, the network needs to ensure that only qualified or authenticated services are available to the vessels or maritime devices, and vice versa, for maritime safety and security. In addition to the multiplicity of services,  the communication infrastructure must connect heterogeneous devices that range from the low-end or low-cost type with reduced functionality to the high-end type with advanced functionality. Low-cost devices are adopted, for example, controlling the surroundings using low-power consumption devices, which generate a small amount of data. The high-end devices are adopted onboard large vessels that encompass dense sensor networks, advanced navigation control systems, and navigation-support systems.\\
In this scenario of both heterogeneous application services and devices, maritime traffic can rapidly grow,  as expected, with the diffusion of IoT-based services such as the remote command and control for autonomous ships or the AI algorithm for the assisted navigation. Hence, the communication infrastructure must be designed considering both scalability needs in terms of bandwidth and computing resources; and capacity constraints due to the maritime hardware in terms of radio spectrum and bandwidth of the channels. So, the communication infrastructure must provide communication and computing support for heterogeneous devices and services. For such a reason, the communication infrastructure must offer interoperability functionalities, i.e., the communication infrastructure must allow the data exchanging and the integration of the information flexibly, effectively, consistently, and cooperatively. Only guaranteeing the interoperability,  it can be provided with access to the network for the different maritime applications and services, based on the IoT paradigm, seamlessly both within and across network boundaries, and provide portability of information efficiently and securely across the complete spectrum of maritime IoT services without effort from the end-user or host, regardless of its manufacturer or origin. The allocation of the radio communication spectrum is the last challenge for the design of the communication infrastructure with a global coverage nature. Indeed, the radio spectrum is typically allocated opportunely by each nation following its regulations. To successfully deploy a communication infrastructure worldwide and to function correctly, it is imperative that an international frequency band is available and established with appropriate international standards and regulations.\\
A communication infrastructure based on hardware components suitable for machine type traffic solves integration between the various types of devices and services relying on the IoT paradigm. Moreover, these technologies can allow the communication of thousands of devices, known as massive-MTC (mMTC) \cite{r46}, \cite{r47}. Concerning the possibility of providing ubiquitous connectivity, the choice to exploit satellite technology offers several advantages: coverage over broad geographical areas, integration with the IoT paradigm as widely reported in the literature, as in \cite{r48}, \cite{r49}, \cite{r49bis} and\cite{r50}. Finally, spectrum allocations for satellite communications in the maritime environment have been already defined, as highlighted in \cite{r51},\cite{r52}, and \cite{r53}.\\
Fig. \ref{fig:comm} shows an example deployment of a space-earth integrated maritime MTC infrastructure, based on the communication modes defined in \cite{r51},\cite{r54}, and \cite{r55}. A mobile station is an MTC terminal aboard a vessel or embedded in marine equipment in this infrastructure. It provides access to the infrastructure for marine equipment or an Ad-hoc network among vessels. We can deploy the maritime cloud in the shore station, acting as a trusted platform to provide various maritime services and applications with the highest computational and storage capacity in the maritime IoT framework. The functionalities for the management and orchestration of the communication infrastructure can run in the cloud. Examples of these functionalities are the dynamic resource allocation, service resolution, and forwarding mechanisms; thus, the physical infrastructure resources can be maximally shared among service providers and are fine-tuned to meet the individual service requirements enabling service-centric networking. A dense network of shore control stations (nearshore) can be deployed, allowing communication among the vessels in the shore and maritime clouds' proximity. This kind of communication can generate more hyper-dense traffic than that originated during offshore communication. Note that vessels can achieve ad-hoc communications to facilitate direct communication among nearby mobile stations for maritime proximity services. In this infrastructure, the satellite can act in both ways providing access to the communication infrastructure or acting as a relay toward a shore station.\\
GEO satellite technology has shown some drawbacks compared to terrestrial radio ones, such as the inefficient covering of a high-dense traffic area due to its large footprint \cite{r56}, high propagation delays, which affect both bandwidth management \cite{r57} and congestion control algorithms \cite{r59}, and error-prone links, which require forward error correction techniques to guarantee service reliability \cite{r58}. Notwithstanding, the new generation of LEO satellites offers a good alternative for maritime \cite{r43} and IoT \cite{r60} communications, as in the scenario taken into account in this work. LEO satellites have a propagation delay of about 10 ms and a smaller footprint than geostationary satellites, so multiple satellites configured in constellations can be used to provide continuous coverage. The satellites in the constellation communicate via inter-satellite links and toward the ground stations through feeder links. Despite the reduced footprint and delay, LEO satellites can be yet unsuitable to serve areas with high-dense traffic, such as harbours and waterways \cite{r56}. In this case, a network of shore stations can allow partitioning the communication effort in the high-dense traffic areas, handling the communications coming from a cluster of vessels.\\
Furthermore, the shore stations can facilitate the optimised assignment of the satellite radio channels, increasing the communications infrastructure's capacity.  In this scenario, tight interference management, i.e., intra-cell and inter-cell co-channel interference, is critical in achieving high spectral efficiency and system capacity in such dense deployment. Note that the shore station could need centralised medium access control to reap the full benefit of the terrestrial communication infrastructure.\\

\begin{table*}
\centering
\caption{Native and Non-Native Maritime Communication Technologies vs. Communication Scenarios.}\label{Table:Comparison}
 \begin{tabular}{ m{1cm}|m{1.8cm}|>{\centering}m{1cm}|>{\centering}m{1.2cm}|>{\centering}m{1cm}|>{\centering}m{1.5cm}|>{\centering}m{1cm}|>{\centering}m{1cm}|m{1cm}}
\multicolumn{2}{c|}{} & \multicolumn{4}{c|}{Native} & \multicolumn{3}{c}{Non-Native} \\
\cline{3-9}
\multicolumn{2}{c|}{} & AIS & SAT-AIS& VDES & SAT-VDES & SatCom & WiFi & 4/5G\\ 
\hline
\multirow{2}{*}{ S2S} & Near-Shore & \xmark & \cmark & \cmark & \xmark & \xmark &\cmark & \cmark \\ 

                       & Off-Shore & \xmark & \cmark & \cmark & \xmark & \xmark & \xmark $\:\backslash$ \cmark &\xmark $\:\backslash$ \cmark \\
                        \hline
\multirow{2}{*}{ S2I} & Near-Shore & \cmark & \cmark & \cmark & \cmark & \cmark & \cmark &\cmark\\
                       & Off-Shore  & \cmark & \xmark & \xmark & \cmark & \cmark & \xmark &\xmark \\ 
                       \hline
\multicolumn{2}{c|}{MTC} &\cmark  & \cmark & \cmark & \cmark & \cmark & \cmark &\cmark\\ 
\hline
\multicolumn{2}{c|}{BC}    &\xmark  & \xmark & \xmark $\:\backslash$ \cmark & \xmark $\:\backslash$ \cmark  & \cmark &\cmark  &\cmark\\ 
\end{tabular} 
\end{table*}

\Figure[t!](topskip=0pt, botskip=0pt, midskip=0pt)[width=0.9\linewidth]{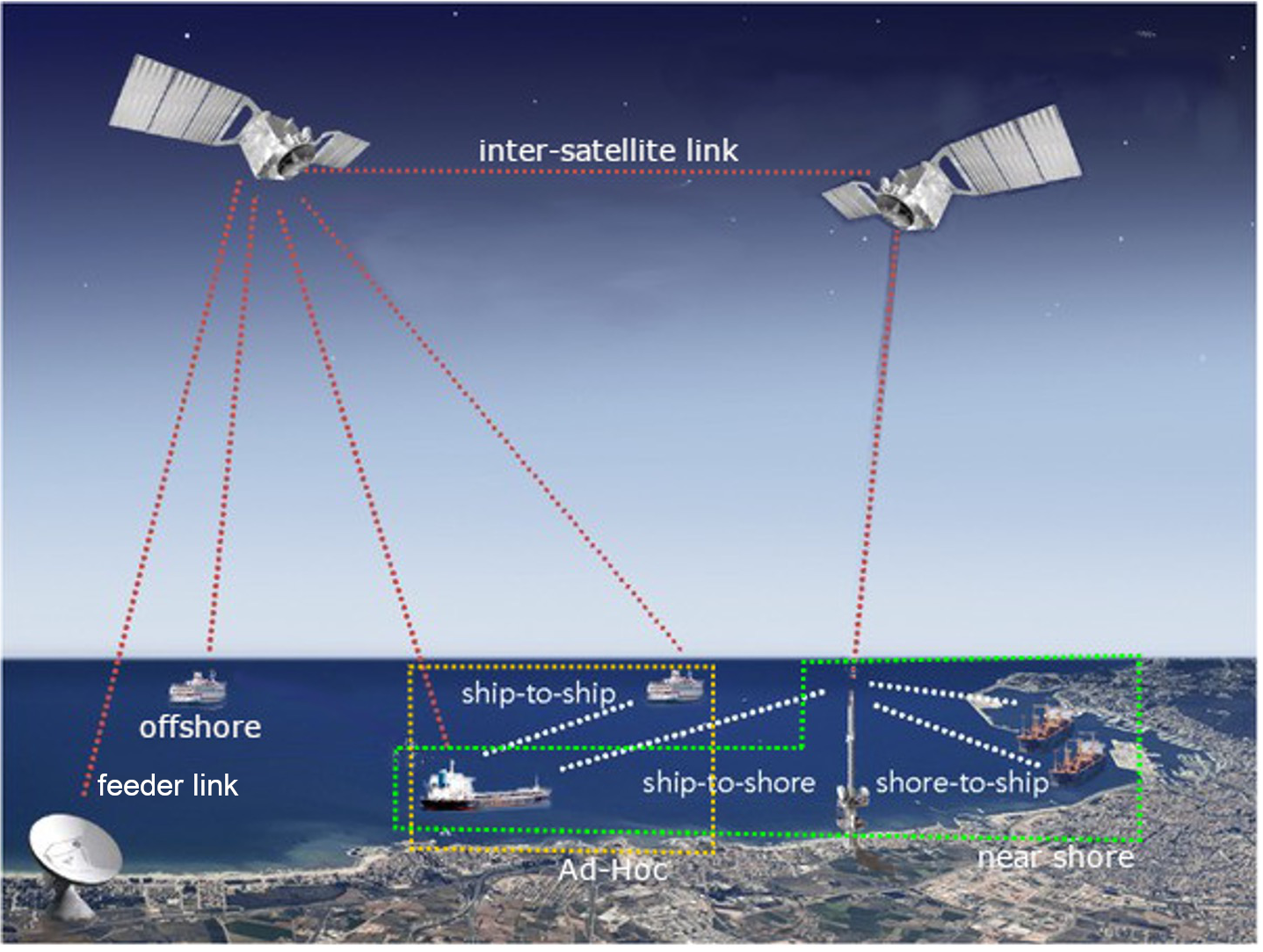}
{Example of Communication Infrastructure.\label{fig:comm}}

Table \ref{Table:Comparison} shows the comparison of the communication technologies considered in this section starting from the existing literature. The first part of the table compares the various communication technologies regarding their ability to provide ship-to-ship (S2S) communication in an ad-hoc way or through ship-to-infrastructure (S2I), for the communication scenarios shown in the Figure \ref{fig:comm}. The first part of the table shows the technology's ability to support communication near the coast or in the harbour (near-shore) or in the open sea (off-shore) where it is not possible to rely on any communication infrastructure. Note that, in the table are evaluated both the WiFi and 4/5G technologies, which offer ease of deployment of both ad-hoc and infrastructure-based communications networks. Both technologies provide data rate levels more than adequate for the applications considered in this paper, but the WiFi has the drawback of the poor communication ranges especially when obstacles are in the communication areas, such as in a harbour. Instead 4/5G has the drawback that the communications always flow through the network infrastructures of the providers. The last consideration should be made about the VDES technology widely used onboard the ships. VDES technology in both satellite (SAT) and terrestrial (TER) versions can provide data rates that are not exceptionally high. However, some studies considered in this section \cite{r51,r52} have highlighted the potential of this technology to be exploited for the transmission of information with data rates close to megahertz.\\
In the second part of the table, the same technologies were compared in terms of their ability to support different levels of data traffic. Precisely, we consider the medium/low levels of data traffic provided by the MTCs, such as that generated by small sensors and actuators of cyber-physical systems of the ship, up to high levels of traffic provided by the Broadband Communications (BC), such as that generated by applications that exploit the control logic based on AI algorithms or applications that exploit the logic of reality and augmented vision.


\section{MODEL-SCALE BENCHMARK}
\label{sec:bench}
It would not be feasible to design, test and validate the system-defined so far due to the cost of applying the proposed idea in a full-scale; a cyber-physical scenario will be the best compromise to merge reliability accuracy. Therefore, a cyber-physical scenario would be more suitable in the preliminary design phase. Such a scenario includes real model scale ships, digital twins, and accurate digital models can simulate real operation conditions. In such a scenario, part of the data needed to test and validate the system has to be provided by the monitoring system installed on ship prototypes (field data), while other data can be simulated or extracted from a database. \\
The scenario should include model-scale vessels, fully actuated and remotely controlled. The model-scale ships should have its digital twin, a digital model that will be updated with the real information coming from the field. Fig. \ref{fig:test} shows that several simulation models of different ship types should interact in this scenario to have a heterogeneous fleet. A set of sensors, such as Lidar, inertial platform, distance gauge, will feed the AI algorithm with real-time information regarding the surrounding environment. To enable the physical communication among the different entities described above, wireless communication interfaces and protocols must be investigated within the envisaged scenarios, for example, through an emulated satellite or a terrestrial long-range wireless link, implemented via software-defined radios.\\
At least three different scenarios should be tested to face the most common encounter situations. The first one includes navigation in blue waters, in which two ships navigate and can communicate with each other. The blue water scenario is used to test the ship-to-ship communication structure, to test and fine tune the local control parameters of the track-keeping system. The second one represents a crowded scenario in a narrow sea, i.e. channels, port entrance, etc, in which the VTS will manage and coordinate the ships' route. The second scenario deals with ''last mile'' navigation operations and is necessary to test the capability of the AI algorithms to handle the navigation of several ships navigating in the same area. This scenario is useful to validate the collision detection and avoidance algorithms. Moreover, in these challenging conditions, the communications infrastructure and the data orchestration systems have to be fully stressed. The third scenario is similar to the previous one, except for the traffic will be not supervised, and the ships communicate with each other. The third suggested test includes not optimal conditions, for instance, to emulate a VTS component's failure, and it will notice the ability of the system to be fault tolerant, for example, continuing to avoid collisions and adequately managing the traffic in the crowed area, acting only through multiple S2S communications.\\

\Figure[b!](topskip=0pt, botskip=0pt, midskip=0pt)[width=0.9\linewidth]{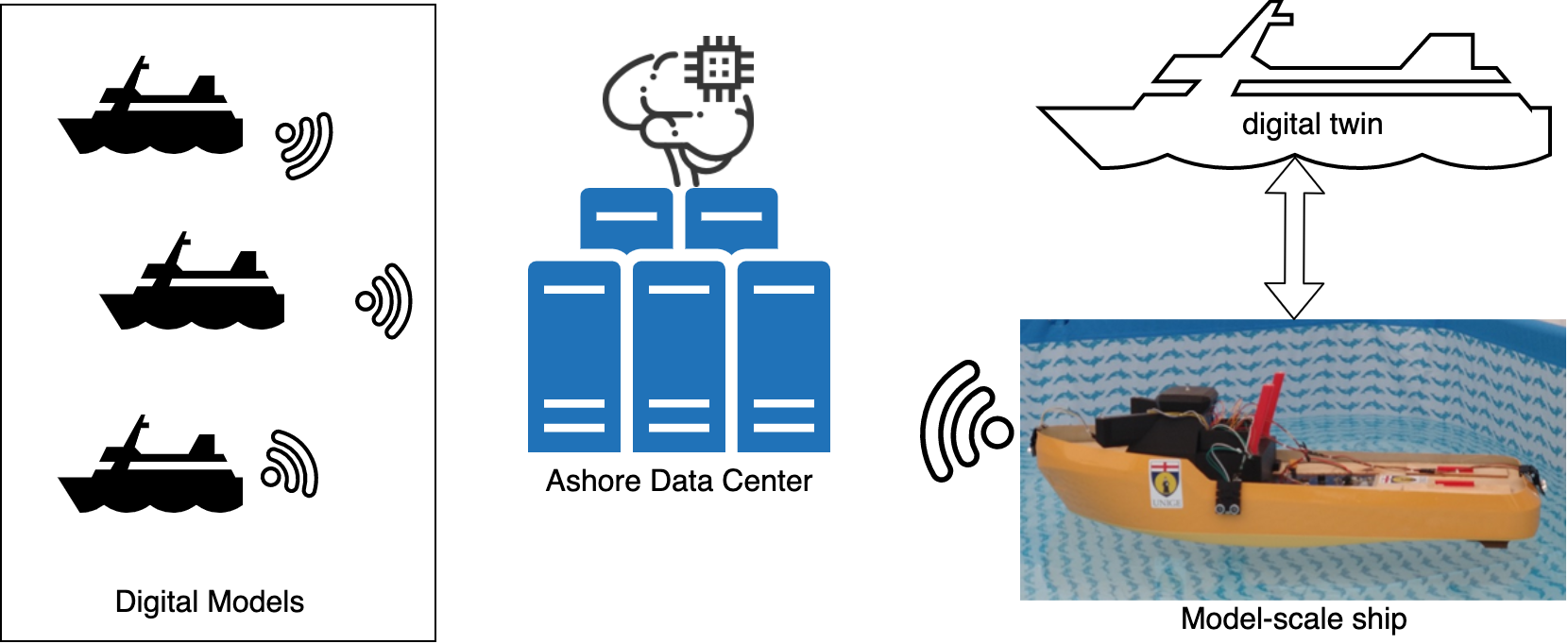}
{Cyber-physical scenario set up.\label{fig:test}}

\section{POTENTIAL BENEFITS AND EXPECTED IMPACT}
\label{sec:benefits}
The proposed system will provide a safe, seamless, smart, inclusive, resilient, environmentally neutral and sustainable system for the maritime sector thanks to ICT technologies and services.\\
The system will boost innovative connected, cooperative and automated sea-mobility technologies and strategies for passengers and goods transportation as additional results.  Automated systems reduce the human factor as a cause of maritime accidents on vessels operating in sensitive areas where maritime accidents and incidents would have a significantly negative impact (coastal zones, marine protected areas). Using an IP-based OCP as a middleware between the GNCs, the ISs will allow a diverse range of equipment or applications to communicate through a mature and standardised environment. This has several advantages: first, it will be possible to integrate commercial or customised IoT sensors and state-of-the-art AI solutions, thus having a potential cost reduction; second, IP connections can be secured using a wide range of existing and validated solutions; third, IP-enabled entities can be allowed to be reached from anywhere, thus paving the way for even more flexible and extendable solutions.  \\
In the same context, the benefits of using deep learning algorithms having predictive capabilities, able to analyse time-varying big data, coupled with innovative displaying tools and technologies allow the reduction of collision and mitigate the associated risk, achieving a global increase of the reliability of the autonomous navigation system.
The foreseen results are summarised hereinafter: 
\begin{itemize}
\item new traffic management system for the coastal area;
increase of the autonomy level of ships, including AI algorithms used on top of GNC systems;
\item improved performance on the collision avoidance systems thanks to new re-planning algorithms for multiple ships; 
\item a fully function cyber-physical scenario that can be used by industry (in particular automation provider and classification bodies) to test and certified new solutions;
enhanced operations and interactions of the autonomous ships with the monitoring operators at the ACC;
\item methodology and tools to acquire and share information from various ships, and to and from the ACC;
\item advanced intelligent methodologies to process data and to achieve safe and reliable automatic navigation;
\item new data representation and displaying techniques that interact with an intelligent system to support the monitoring operators of ashore centres;
\item an efficient OCP to dynamically adapt the system to the different operational situations while satisfying Quality of Service, Safety and reliability requirements for computing and communication.
\end{itemize}

\begin{figure}
    \centering
    \includegraphics[trim=0 0 0 0, clip=true, width=0.99\columnwidth]{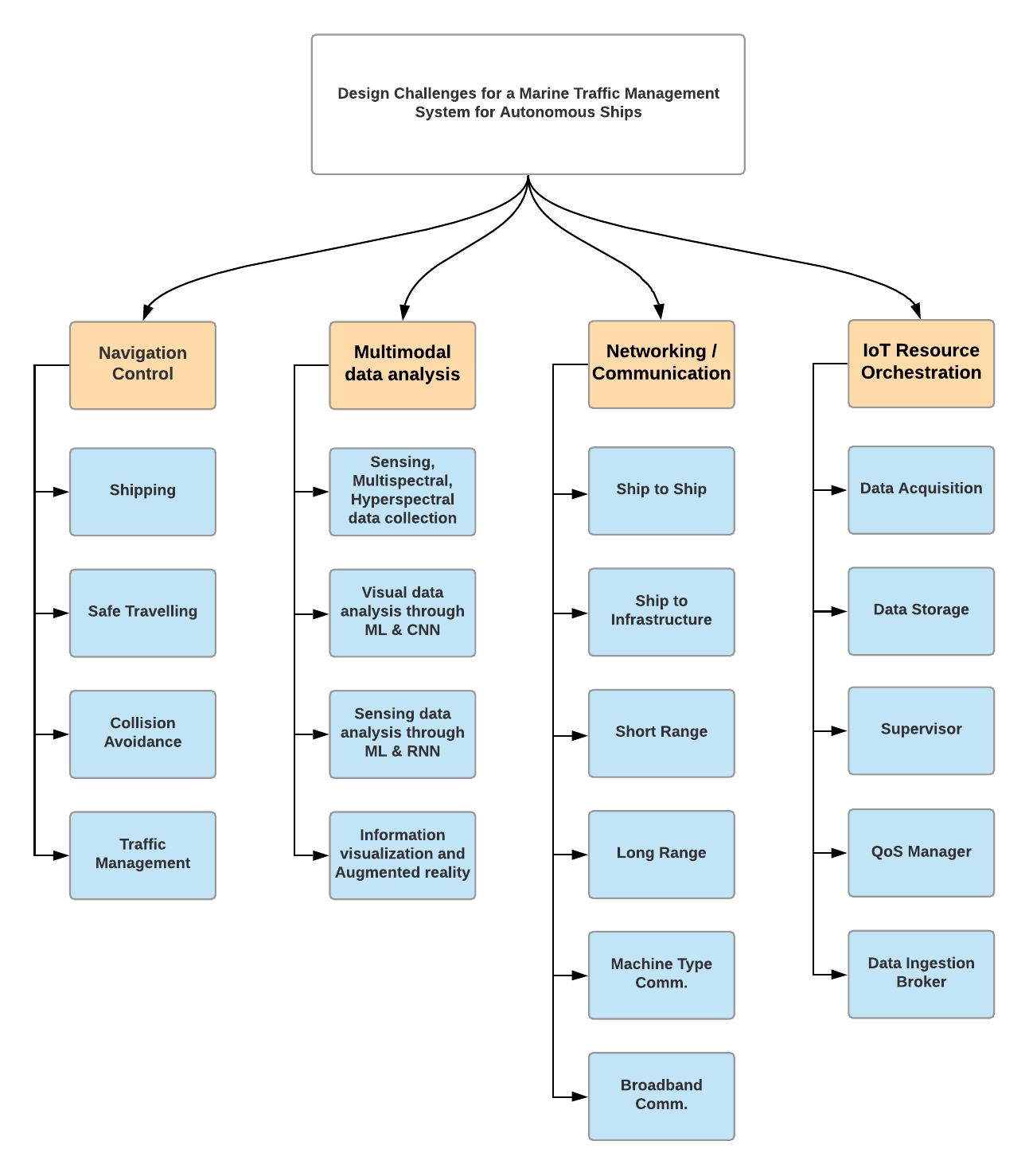}
    \caption{A taxonomy of the challenges to design a marine traffic management system for autonomous ships.}
    \label{fig:taxonomy}
\end{figure}

Figure \ref{fig:taxonomy} shows a taxonomy of the existing challenges to design a marine traffic management system accounting for autonomous ships.
The picture both summarizes and opens to main research blocks that may be involved in the development of a project devoted to this aim. More than this, some peculiar study tasks provide the structure of what still needs investigation, in line with what is presented in the literature review of this article, which stands out as a widespread but not yet complete overview.

Regarding the impact that a project of this magnitude could have on the induced world of work, there can be no doubt that replacing crew with autonomous technology will severely impact today's type of seafarers. Seafarers work in a hazardous and stressful environment far from home, the reason why few sailors have an entire career at sea. 

Autonomous ships imply that jobs onboard will be replaced by alternative jobs in a better work environment onshore. An evident conclusion is that the profession of sea officer is unattractive to women and that both women and men do not aspire to a whole career at sea. The transition of jobs from sea to Back Office environments will improve the working conditions of personnel employed in shipping and their social and personal life, will be more attractive to women and contribute to gender equality in the shipping sector with job opportunities and enhanced economic participation.


\section{Conclusions}
\label{sec:conclusion}
From a general perspective, the proposed system will directly impact the maritime field, thanks to its potential of bringing a concept of dynamic cooperation in the scenario of autonomous ships. In such a sector, the idea is to enhance the existing technology using augmented and virtual reality and machine learning approaches to obtain information suitable for the reliable control of autonomous surface vessels in a wide range of navigation scenarios.\\
Regarding the use of AI algorithms in the challenging maritime context, improvement needs to be done to achieve the most effective and efficient way of extracting information and representing it concisely and clearly through augmented-reality devices. For what concerns; instead, the orchestration and the communication techniques should be designed specifically for the maritime context, fitting the challenging environmental conditions and meeting the stringent safety requirements. \\
For the maritime sector, the proposed system will be a breakthrough in the current state of the art, looking forward to when several ships, with different control strategies developed by different automation providers, will navigate together with crewed ships. In such a scenario, Artificial Intelligence algorithms, ship to ship communications with a standard protocol, enhanced concerning the Automatic Identification System currently in use, will be the solution for the new generation of sea traffic management. To the author's best knowledge, nothing similar exists yet, neither in scientific literature nor in industrial projects.
In the maritime sector, artificial intelligence has begun to establish itself in recent years due to a growing demand for ships' autonomy. \\
Adopting automatic technologies that support the service management will allow rationalising the assets (e.g. energy, materials, equipment) involved in the automated services. This rationalisation benefits both providers and society in terms of product price and environmental pollution. Moreover, the automated technologies that allow automated service behaviour analysis can be adapted to support the workers by predicting the hazardous and stressful conditions. 
Moreover, allowing automated interactions among multiple entities (ships and shore) will ease the adoption of autonomous ships, simplifying diverse systems' interactions while ensuring optimal performance.\\
Autonomous ship capabilities will impact both the ship investment costs (CAPEX) and the ship operational costs (OPEX). Crewless ships will not need deck houses, heating/cooling and lifesaving equipment. This will result in different ship concepts with more space for payload, lower CAPEX and lower carbon footprint per cargo unit than current ship designs. On the other hand, the cost of autonomous ship technology will be higher than present-day ship automation. Preliminary estimates indicate those cost-savings due to the absence of ship personnel are balanced by the ship automation's higher cost. The proposed system will decrease the number of collisions (ship to ship and ship to the pier), with considerable repair costs saving and reducing the delay in goods/services.\\
Finally, the proposed system will increase the safety of human life and goods at sea, reducing accidents in the coastal area. Secondarily, the reduced risk of accidents automatically reduces the oil spills significantly, with a lower impact on the nowadays fragile marine ecosystems. 

\section*{ACRONYMS}
\begin{center}
\begin{tabular}{ll}
ACC	&	Ashore Control Center\\
AI 	&	Artificial Intelligence\\
AIS & Automatic Identification System\\
AR	&	Augmented Reality\\
ASV &		Autonomous Surface Vessels\\
BC & Broadband Communication\\
CoAP &		Constrained Application Protocol\\
DL	&	Deep Learning\\
GNC &		Guidance-Navigation-Control\\
ICT	&	Information and Communication Technologies\\
IoT &		Internet of Things\\
IS 	&	Intelligent System\\
ML	&	Machine Learning\\
MTC &		Machine-Type Communication\\
OCP &		Orchestration and Communication Platform\\
QoS &		Quality of Services\\
VDES&  VHF Data Exchange Data System\\
VR	&	Virtual Reality\\
VTS &		Vessel Traffic System\\
\end{tabular}
\end{center}
\bibliographystyle{IEEEtran}
\bibliography{bib1}{}
\vskip -2\baselineskip plus -1fill
\begin{IEEEbiography}[{\includegraphics[width=1in,height=1.25in,clip,keepaspectratio]{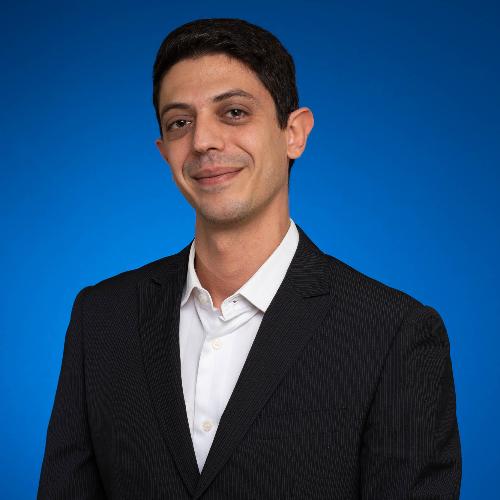}}]{Michele Martelli} was born in Italy in 1985. He received the BS, MS, and PhD degrees in marine engineering and naval architecture in 2006, 2009, and 2013, respectively. Currently, he is an Associate Professor at the University of Genoa. He is a co-author of more than 60 scientific papers. His main research interests focus on the study of the dynamics of the propulsion plant and its control system, and autonomous navigation. Since 2010, he has been actively working on several research projects, funded by both private and public companies.
\end{IEEEbiography}

\begin{IEEEbiography}[{\includegraphics[width=1in,height=1.25in,clip,keepaspectratio]{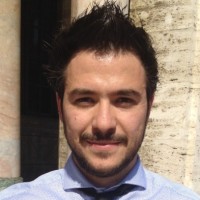}}]{Antonio Virdis}  is Assistant Professor at the University of Pisa, where he obtained his MSc degree in Computer System Engineering in 2011, and his PhD in Information Engineering in 2015. His research interests include Quality of Service, Edge Computing, network simulation and performance evaluation. He coauthored more than 60 peer-reviewed papers, 8 patents and 4 book chapters in the above fields. He edited the book tile "Recent Advances in Network Simulation", published by Springer/EAI. He is and has been involved in research projects supported by private industries and funded by the EU community. He served as TPC chair for the International OMNeT++ Summit, and for the IEEE SmartSys workshop, and as a member of the TPC for more than 20 international conferences. He is one of the author and maintainers of the SimuLTE and Simu5G open-source projects, for the system-level simulation of 4G and 5G communication networks. 
\end{IEEEbiography}

\vskip -2\baselineskip plus -1fill
\begin{IEEEbiography}[{\includegraphics[width=1in,height=1.25in,clip,keepaspectratio]{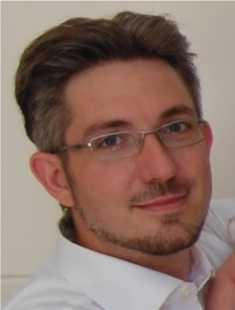}}]{Alberto Gotta}  received his M.Sc and Ph.D in 2002 and 2007, respectively, he is a researcher at the Wireless Networks Laboratory at CNR-ISTI, the Institute of Information Science and Technologies of the National Research Council (CNR), Italy. He is member of the IEEE International Network Generations Roadmap (INGR) Satellite Working Group. His expertise is mainly related to traffic engineering applied to satellite, machine-to-machine, and UAV networks. He has participated and leaded several EU, National, and ESA funded R\&D projects. He co-authored more than 80 papers and has served the TPCs of flagship ComSoc conferences and Symposia.
\end{IEEEbiography}

\begin{IEEEbiography}[{\includegraphics[width=1in,height=1.25in,clip,keepaspectratio]{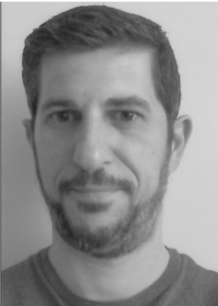}}]{Pietro Cassarà} received his M.Sc. degrees in Telecommunication and Electronic Engineering from University of Palermo in the 2005, and its Ph.D. degree in the 2010, jointly with the State University of New York. Nowadays, he is staff member of the Institute of Science and Information Technologies (ISTI), at the National Research Council (CNR), Pisa, Italy, and since 2017 he has been temporary staff member of the CMRE lab at the NATO of La Spezia. He is currently a member of the IEEE ComSoc and VTS Committees, his research interests include wireless sensor network and IoT communications. He has been participating in European, ESA, and National funded projects.
\end{IEEEbiography}

\begin{IEEEbiography}[{\includegraphics[width=1in,height=1.25in,clip,keepaspectratio]{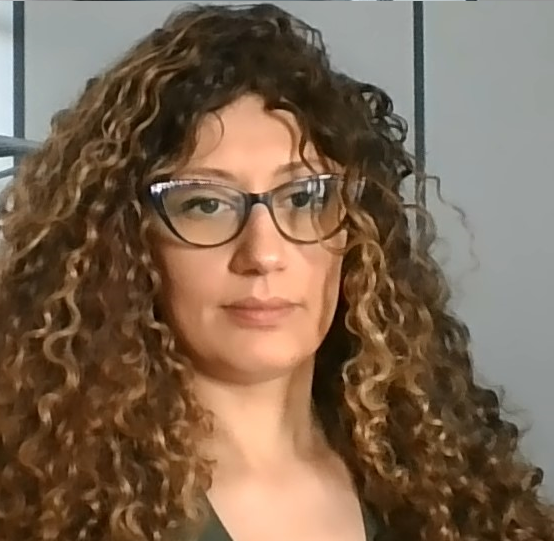}}]{Maria Di Summa} is researcher at STIIMA-CNR since 2010. She completed her Ph.D. and her undergraduate studies at Politecnico di Bari. Her research interests includes Virtual, Augmented and Mixed Reality, People and posture tracking, Intelligent Systems, AI paradigms for behaviour learning and understanding, feature extraction for pattern recognition, Anomalies detection. She served as Industrial Session Chair for EuroVR 2016. She has collaborated, covering different roles, actively in several Regional, National and International projects 
focused on Advanced Manufacturing and Ambient Assisted Living.
\end{IEEEbiography}
\vskip -2\baselineskip plus -1fill
\EOD
\end{document}